\documentclass[conference]{IEEEtran}
\IEEEoverridecommandlockouts
\usepackage{cite}
\usepackage{amsmath,amssymb,amsfonts}
\usepackage{algorithmic}
\usepackage{graphicx}
\usepackage{textcomp}
\usepackage{xcolor}
\usepackage{epsfig}
\usepackage{url} 
\def\BibTeX{{\rm B\kern-.05em{\sc i\kern-.025em b}\kern-.08em
    T\kern-.1667em\lower.7ex\hbox{E}\kern-.125emX}}
\begin{document}

\title {Advancements in Continuous Glucose Monitoring: Integrating Deep Learning and ECG Signal}

\author{\IEEEauthorblockN{MohammadReza Hosseinzadehketilateh, Banafsheh Adami, Nima Karimian}
\IEEEauthorblockA{\textit{Lane Department of Computer Science and Electrical Engineering} \\
\textit{West Virginia University}\\
Morgantown, USA \\
mh00136@mix.wvu.edu, ba00011@mix.wvu.edu, and nima.karimian@mail.wvu.edu}
}

\maketitle

\begin{abstract}
This paper presents a novel approach to non-invasive hyperglycemia monitoring utilizing electrocardiograms (ECG) from an extensive database comprising 1119 subjects. Previous research on hyperglycemia or glucose detection using ECG has been constrained by challenges related to generalization and scalability, primarily due to using all subjects' ECG in training without considering unseen subjects—a critical factor for developing methods with effective generalization. We designed a deep neural network model capable of identifying significant features across various spatial locations and examining the interdependencies among different features within each convolutional layer. To expedite processing speed, we segment the ECG of each user to isolate one heartbeat or one cycle of the ECG. Our model was trained using data from 727 subjects, while 168 were used for validation. The testing phase involved 224 unseen subjects, with a dataset consisting of 9,000 segments. The result indicates that the proposed algorithm effectively detects hyperglycemia with a 91.60\% area under the curve (AUC), 81.05\% sensitivity, and 85.54\% specificity.
\end{abstract}

\begin{IEEEkeywords}
ECG, hyperglycemia, Deep learning
\end{IEEEkeywords}

\section{Introduction}
Continuous monitoring of hyperglycemia is crucial for individuals with diabetes. It enables both individuals and healthcare providers to make real-time adjustments to medication and insulin doses. Continuous glucose level monitoring helps maintain blood glucose levels within the desired range, reducing the risk of complications. According to CDC~\cite{CDC}, 38.4 million people in the US have diabetes, and 97.6 million people aged 18 years or older have prediabetes. Thus, developing continuous monitoring is essential. Traditional methods for monitoring blood glucose include finger-prick measurements and continuous glucose monitoring (CGM) devices~\cite{villena2019progress}. However, finger-prick measurements are associated with pain, discomfort, and high costs. Additionally, this approach fails to offer a continuous monitoring solution for blood glucose levels. Recent studies have explored the use of electrocardiograms for hyperglycemia detection, making them a viable option for continuous monitoring. This data can be conveniently collected from wearable devices like the Apple Watch or Fitbit. In contrast to alternative, noninvasive methodologies, the acquisition of ECG signals through wearable devices is readily achievable. This presents a cost-effective and comfortable monitoring solution for individuals with diabetes~\cite{lazaro2020wearable,sethuraman2021mywear}.

Ngyuen et al.~\cite{nguyen2012identification} demonstrated that ECG exhibits notable changes during hyperglycemia events. The study, involving five subjects who experienced both hyperglycemia and hypoglycemia, revealed significant increases in heart rate, QTc (corrected ECG QT interval), and RTc (corrected ECG RT interval). Subsequently, a neural network was developed, utilizing 16 handcrafted features with a dataset of 10 subjects for hyperglycemia identification~\cite{nguyen2014neural}. The proposed model achieved a sensitivity of 70.6\% and specificity of 65.4\%. Li et al.~\cite{li2021non} proposed density-based spatial clustering applications with noise and convolutional neural networks (DBSCAN-CNN), for glucose level identification using ECG data from 21 subjects. The proposed model demonstrated impressive classification accuracy, achieving 87.94\% for low glucose levels, 69.36\% for moderate glucose levels, and 86.39\% for high glucose levels. Furthermore, the reported sensitivity and specificity were 98.48\% and 76.75\%, respectively. Following Li et al.'s work~\cite{li2023noninvasive}, the ResNet architecture was employed, yielding similar results. While prior research has conducted preliminary investigations into the potential utilization of ECG for detecting blood glucose and hyperglycemia, the existing studies not only lack comprehensive demonstrations of high accuracy and specificity but also suffer from limited data, typically involving a maximum of 21 subjects. Our goal is to address the aforementioned challenges and establish the feasibility of ECG for blood glucose level detection or hyperglycemia by introducing a novel database comprising 1116 subjects. Additionally, we propose a new deep learning architecture that surpasses the performance of existing methodologies even when confronted with a larger dataset.

The main contributions of this paper are summarized:
\begin{itemize}
    \item We propose a novel and generalized convolutional block attention module with CNN for hyperglacemia detection using a non-invasive ECG signal. 
    \item  We present a large ECG database consisting of 1,119 subjects, where each subject is equally labeled for both high levels of hyperglycemia and non-hyperglycemia.
    \item  We evaluated the performance of hyperglycemia detection in diverse scenarios, including situations involving unseen subjects and individual-based analyses.
    \item To showcase the effectiveness, robustness, and generalization capability of our proposed methods, we conducted experiments on a novel ECG database comprising 68,274 samples recorded from 1119 subjects with a sensitivity of 81.05\% and a specificity of 85.54\%.
    \item Our proposed work achieved a significant improvement compared to results in~\cite{li2023noninvasive, cordeiro2021hyperglycemia}. Our proposed model improved the specificity by 10\%.
    
\end{itemize}

\section{Experiment Setup}
\subsection{Data Acquisition}
The experiments were meticulously conducted in controlled settings to mitigate potential external interference. Prior to the beginning of the experiment, participants did not engage in any physical exercise or drug consumption. ECG signals were captured using an Analog AD-8232 with a sampling rate of 1000 Hz, and blood glucose values were acquired from participants' fingers using a minimally invasive BG meter (ACCU-CHEK). A total of 1119 subjects, including 386 females and 733 males aged between 38 and 80 years, participated in the study. Fasting was not mandatory, and participants did not disclose their overall health status. Each participant participated in two successive recording sessions, both conducted in the morning. Each session consisted of the recording of a 60-second single-lead ECG and blood glucose concentration.Following this, a comprehensive analysis was conducted on each ECG recording, leading to the exclusion of those with low quality, resulting in a dataset comprising 1963 recordings. Recordings from subjects with glucose concentrations exceeding 100 mg/dL were identified as characteristic of hyperglycemia for the purposes of this study.

\subsection{Data Processing}
As the ECG encompasses diverse noise sources such as baseline wander (BW), motion artifact (MA), and electrode movement (EM), these can improve the accuracy of proposed model by smoothing the ECG signal. To mitigate potential artifacts arising from the setup and removal of electrodes on the subject, the initial and final 2 seconds of the raw ECG signals are disregarded. The remaining data undergoes filtration employing a Butterworth bandpass filter of order 4, with a frequency range set at 1 Hz to 40 Hz~\cite{ingale2020ecg}. 

ECG waveforms comprise a repetitive order characterized by five major peaks, namely P, QRS, and T. Given that each ECG heartbeat contains redundant information, repetitively processing heartbeats with a correlation is inefficient. ECG segmentation emerges as a widely adopted signal preparation method to reduce signal size for subsequent feature extraction. Essentially, the goal of segmentation is to identify recurring patterns in the ECG signal, specifically the P, QRS, and T waves, thereby significantly reducing the input size of deep learning. In our study, segmentation is aligned with the reference point, typically identifying an R-peak, and fixed distances before and after the identified R-peaks. This involves extracting a partial ECG signal (R-$t_{1}$, R+$t_{0}$) instead of the entire signal, where $t_{1}$ and $t_{0}$) represent predefined fixed times covering the majority of the P-QRS-T fragment. Subsequently, we illustrate the influence of using a single segment of the ECG signal as input for deep learning, as well as employing five consecutive subsequences of the ECG signal to assess our base model.  Segments across subjects do not share the same unit and require normalization to standardize them. This process involves not only removing their mean but also scaling them to achieve unit variance before feeding them into the deep learning model. The normalization calculation, utilizing the mean and standard deviation, was performed using the StandardScaler function available in the Python Scikit-learn library.

\begin{figure}
  \centering
  \begin{tabular}{cc}
    \includegraphics[width=0.48\linewidth]{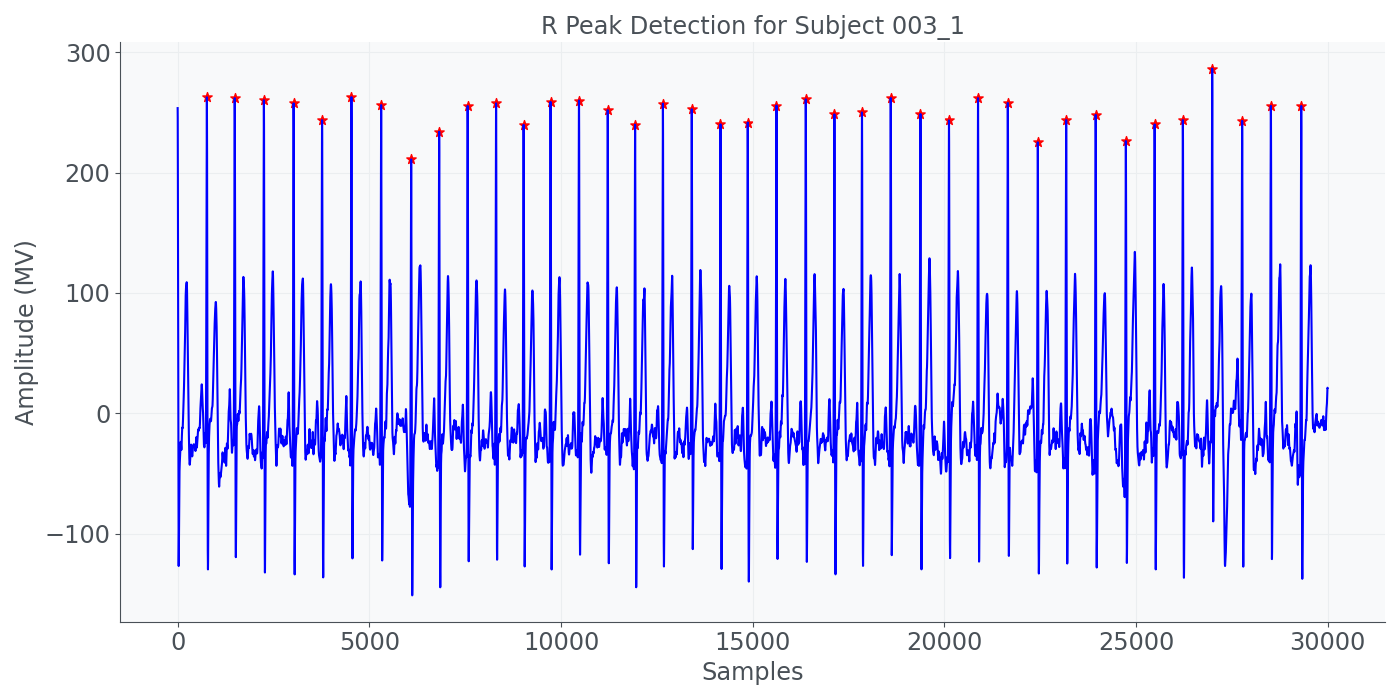} &
    \includegraphics[width=0.48\linewidth]{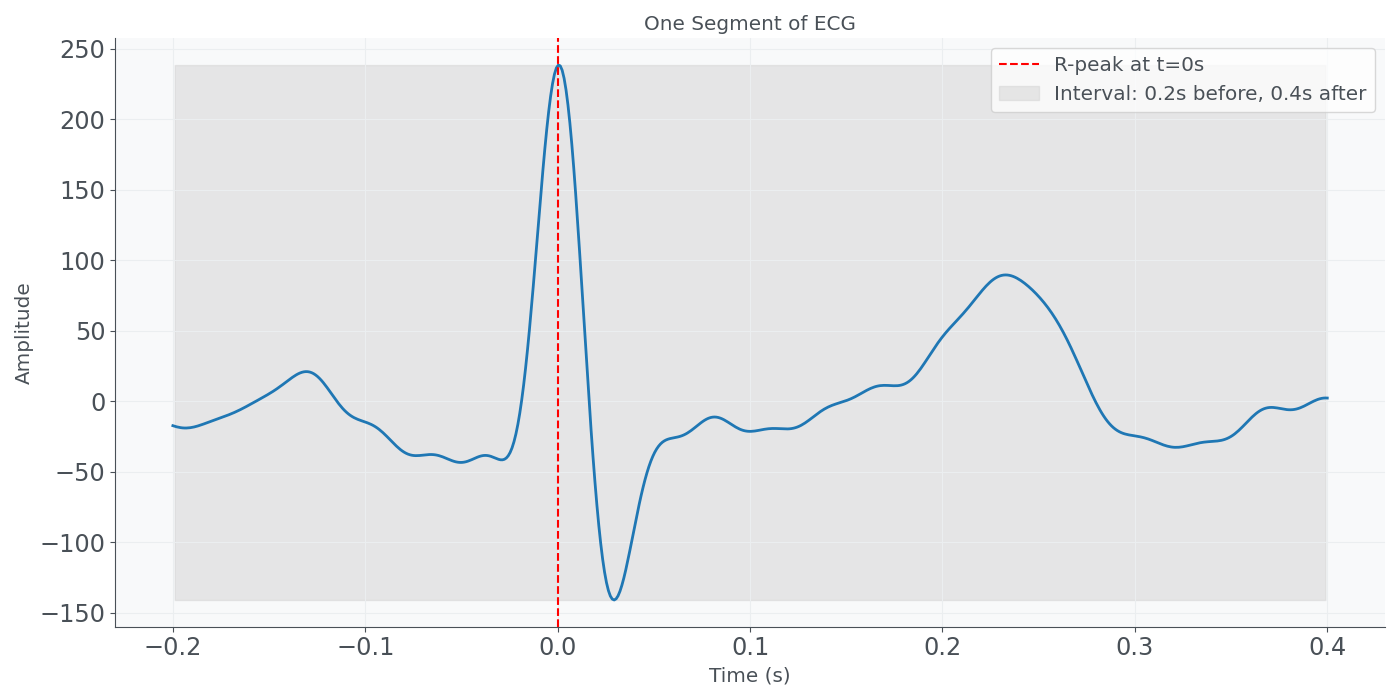} \\
    (a) & (b)\\
  \end{tabular}
  \caption{(a) Demonstrate a recording of ECG waveform from a participant. (b) Shows one segment of ECG signal.)}
  \label{fig:ecgsegment}
\end{figure}

\begin{figure*}
    \centering
    \includegraphics[width=0.9\linewidth]{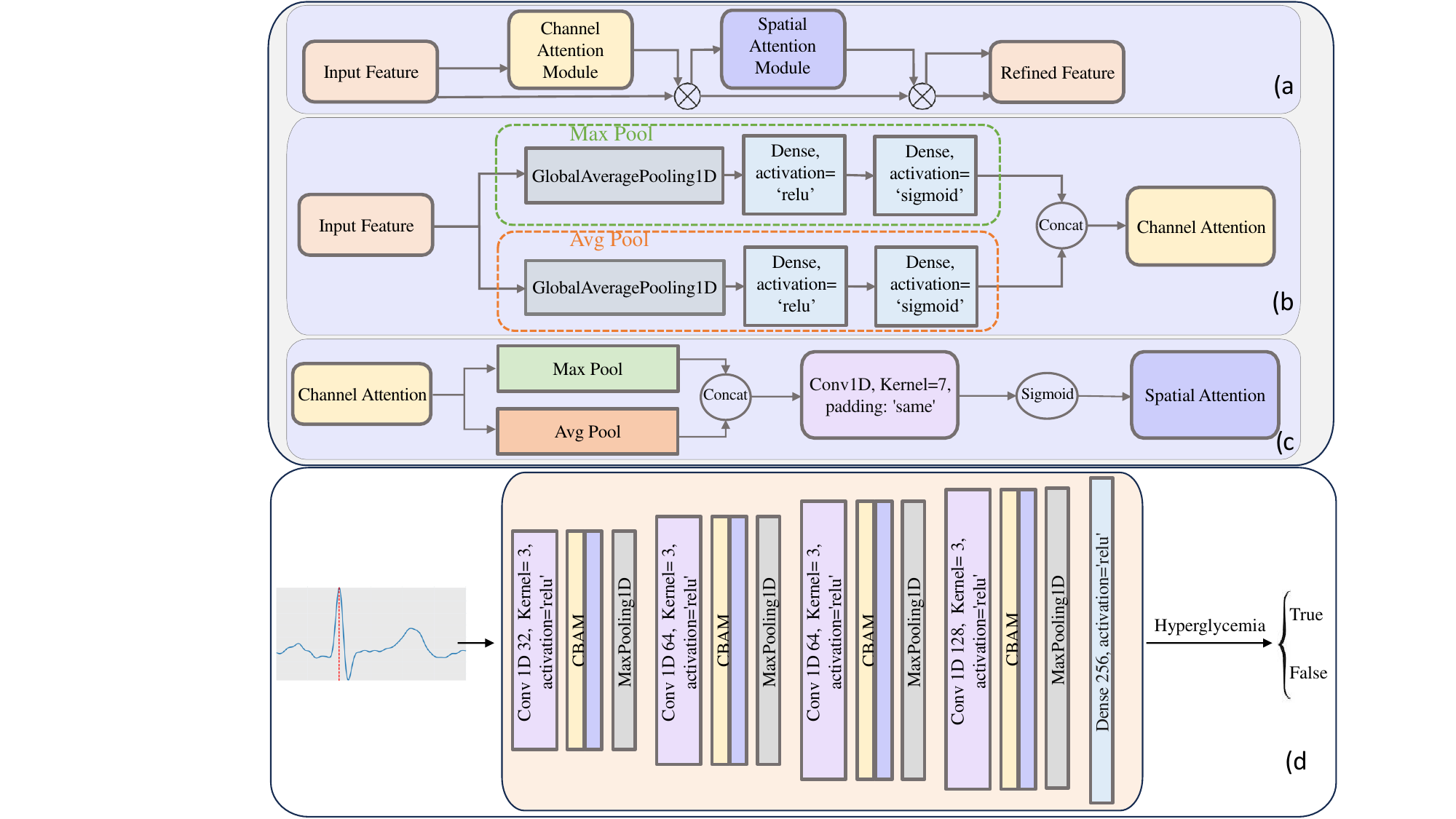}    
    \caption{proposed hyperglacemia detection using CNN with block attention module. a) convolutional block attention module (CBAM). b) channel attention module (CAM), c) spatial attention module (SAM), d) convolutional neural network with attention mechanism.}
    \label{fig:model}
\end{figure*}

\subsection{Evaluation Metrics}
Three performance metrics were employed to assess the hyperglycemia detection performance: true positive rate (TPR), false positive rate (FPR), and area under the curve (AUC). FPR, or (1-Specificity), represents the percentage of healthy ECG wrongly classified as hyperglycemia, while TPR, or sensitivity, signifies the percentage of hyperglycemic events successfully identified as hyperglycemia. The AUC of the receiver operating characteristic (ROC) curve served as the performance metric for model comparison. In binary classification problems like these, the threshold used to distinguish between the two output labels (hyperglycemia and non-hyperglycemia) directly impacts performance metrics. The ROC curve illustrates model performance in terms of TPR versus FPR across different thresholds, and the area under these curves offers a comprehensive performance measurement for all thresholds. Sensitivity and specificity were also employed as additional metrics for classification performance. Sensitivity gauges how accurately a test produces a positive result for individuals with the hyperglycemia condition, while specificity assesses a test's ability to accurately generate a negative result for individuals without hyperglycemia. 727 subjects with around 29,080 segments are used for training, while 168 subjects, contributing 6,720 segments, are allocated for validation. The testing phase involved 224 unseen subjects, comprising a dataset of 9,000 segments.

\subsection{Model}
Our proposed model is demonstrated in Figure.~\ref{fig:model}. It comprises four layers of a convolutional neural network (CNN), followed by a convolutional block attention module (CBAM) and max pooling~\cite{woo2018cbam, adami2023contactless}. Our proposed CBAM incorporates two attention mechanisms: the channel attention module (CAM) and the spatial attention module (SAM). The channel attention module, as shown in Figure~\ref{fig:model}-b. The channel attention module captures the interdependencies among feature channels in the ECG segment through max pooling and average pooling operations across two dense layers. The first dense layer, utilizing a ReLU activation function, reduces the number of information channels, and the second layer generates channel-wise attention using a sigmoid activation function.

The Spatial Attention Module, illustrated in Figure~\ref{fig:model}-c, focuses on capturing spatial relationships within feature maps of both hyperglycemia and non-hyperglycemia representations in the ECG signal. This enhancement aims to boost the model's performance in classification tasks, enabling it to concentrate on informative spatial regions within the feature maps. The module calculates spatial attention for hyperglycemia input features by performing average-pooling and max-pooling operations along the channel dimension of the hyperglycemia feature map. This process generates a concise and informative feature representation, which is then concatenated. Following concatenation, convolution layers ($f^{7\times7}$) are applied to construct a spatial attention map ($M_{s}\in \mathbf{R}^{C\times H \times W}$), where $C$ is number of channel which we set it to 1, $H$ is height, and $W$ is width. using Equation \ref{eq:spatial}:

\begin{align}
    M_{spatial}(X)=\sigma (f^{7\times7} [AvgPool(X); MaxPool(X)]),
    \label{eq:spatial}
\end{align}

Where $\sigma$ is sigmoid function, $f^{7\times7}$ is convolution operation with filter size $7\times7$. In our study, we fixed the values of Channel ($C$) and height ($H$) at one due to the sequential nature of the signal. Additionally, $W$  was set to 600, representing the size of a single segment in the ECG signal.

The channel attention module focuses on understanding the relationships or dependencies that exist between different feature channels in the ECG signal. 
\begin{align}
        M_{Channel}(x) = \sigma (Maxpool(x)) + \sigma (Avgpool(x))
        \label{eq:channel}
\end{align}
Where $x$ is the original feature map produced by a convolutional layer
This integration allows for the capture of both channel-wise and spatial-wise attention, facilitating the learning of representations for hyperglycemia levels in the ECG signal.

The proposed method tries to minimize the binary cross entropy (BCE) loss, which measures the difference between the predicted probability of hyperglycemia and the actual binary labels (indicating the presence or absence of hyperglycemia). Given an ECG signal as input and the predicted probability $p$ that hyperglycemia is present, along with the true label y (which is 1 if hyperglycemia is present and 0 if not), the BCE loss is calculated as:
\begin{align}
\text{BCE}(y, p) = -[y \log(p) + (1 - y) \log(1 - p)].
\end{align}

\begin{figure}
  \centering
  \begin{tabular}{cc}
    \includegraphics[width=0.48\linewidth]{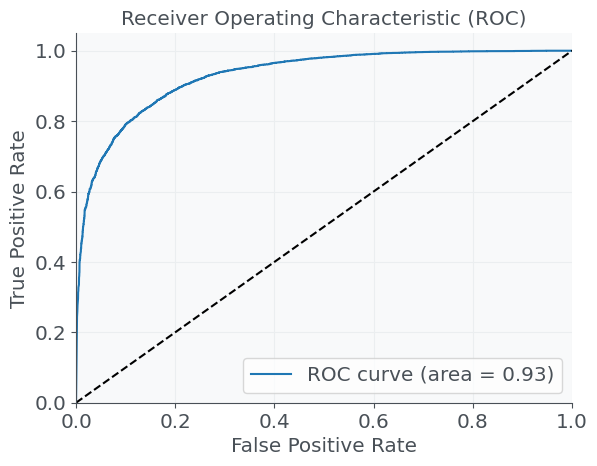} &
    \includegraphics[width=0.4\linewidth]{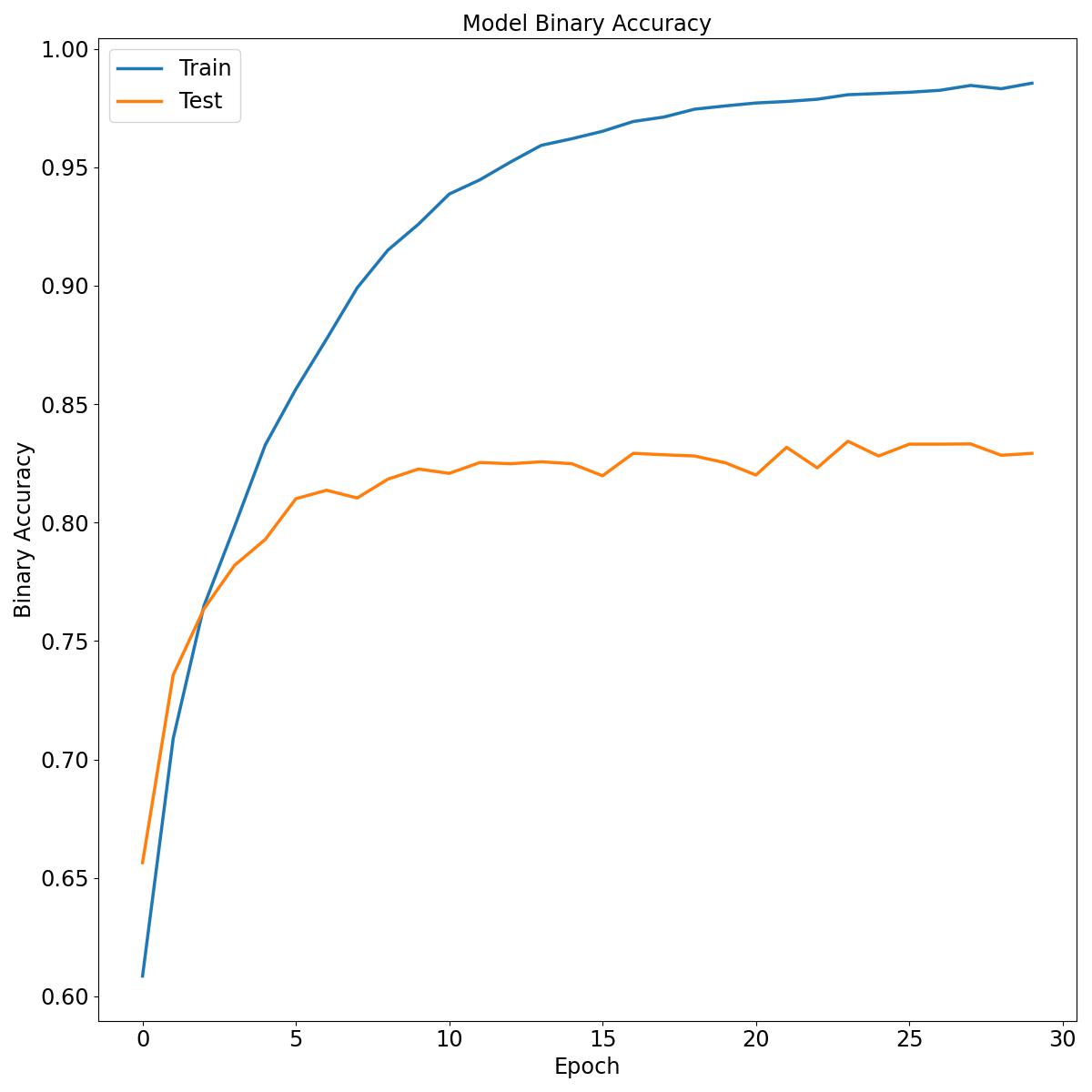} \\
    (a) & (b)\\
  \end{tabular}
\caption{(a) ROC for testing dataset with AUC of 93\%. (b) Depicts model accuracy for both training and testing under different epoch.)}
  \label{fig:result}
\end{figure}

\section{Experimental Results}
According to the results depicted in Figure~\ref{fig:result}, our proposed model, utilizing data from 1119 subjects, achieved notable performance metrics. Specifically, it attained a 91.60\% area under the curve (AUC), 81.05\% sensitivity, and 85.54\% specificity. This evaluation was conducted by training the model on a subset of subjects' ECG data and validating it on unseen subjects. In contrast to other approaches that involve using only a portion of subjects' ECG for training, our methodology reflects a more realistic scenario. Furthermore, unlike previous studies that considered some ECG segments as outliers and removed them, our study retained all segments and subjects, even those with noisy data. This contributes to the robustness and real-world applicability of our proposed model.

\section{Discussion and comparison}
In our previous investigations\cite{cordeiro2021hyperglycemia}, we segmented the ECG data of each participant into approximately 40 consecutive cardiac cycles, where each segment represented a complete ECG waveform corresponding to a cardiac cycle. Subsequently, we applied a random partitioning method, designating 85\% of the segments for the training subset and allocating the remaining as the testing subset. Following this, the training and testing subsets were amalgamated to create the respective training and testing datasets. The choice to employ a segmentation method based on individual ECG data rather than participant-wise stems from the personalized variations observed in both ECG and blood glucose. While our earlier study shared similarities with Li et al.~\cite{li2023noninvasive}, it is not realistic. The primary objective of our current study is to generalize the model based on the current participants, enabling it to identify hyperglycemia in unseen subjects. For instance, if an Apple Watch incorporates this feature, the system should be trained on an existing database and possess the capability to notify a user about hyperglycemia without requiring their data under both conditions. In order to compare our proposed model with the literature, we have implemented the same scenario for training and testing the model, splitting the ECG segments of each subject into training and testing with our proposed deep learning architecture. The model performance is demonstrated in Table~\ref{tab:comparison}. As can be seen in this table, our proposed model outperforms other methods when compared with other methods, and it has been evaluated on more subjects.

\begin{table}[]
\begin{tabular}{l|l|l|l|l|}
\cline{2-5}
 & Sensitivity & Specificity & AUC & \#subjects \\ \hline
\multicolumn{1}{|l|}{Linh et al.\cite{nguyen2014neural}} & 65.64\% & 56.21\% & 61.68\% & 10 \\ \hline
\multicolumn{1}{|l|}{cordeiro et al.\cite{cordeiro2021hyperglycemia}} & 87.57\% & 85.04\% & 94.53\% & 1119 \\ \hline
\multicolumn{1}{|l|}{Li et al.\cite{li2023noninvasive}} & 98.4 & 76.75 & \multicolumn{1}{c|}{-} & 21 \\ \hline
\multicolumn{1}{|l|}{\textbf{Our work}} & \textbf{96.07} & \textbf{95.46} & \textbf{99} & \textbf{1119} \\ \hline
\end{tabular}
\caption{Evaluating the proposed model by comparing it with various frameworks using ECG data from all subjects in training. Measurement will be based on sensitivity, specificity, and AUC.}
\label{tab:comparison}
\end{table}

\section{Conclusion}

In this study, we introduce an innovative approach for non-invasive hyperglycemia monitoring using electrocardiograms (ECG) from a remarkably large database of 1119 subjects. Our methodology involves developing a generalized model trained on a subset of subjects' ECG data, facilitating hyperglycemia prediction in unseen subjects. We developed a deep neural network model capable of identifying significant features across various spatial locations and examining interdependencies within each convolutional layer. Our model underwent training with data from 727 subjects, encompassing around 29,080 segments, while 168 subjects, contributing 6,720 segments, were allocated for validation. The testing phase involved 224 subjects, comprising a dataset of 9,000 segments. We ensured an equal distribution of data containing hyperglycemia and normal instances in both the training and testing sets. The results demonstrate the effectiveness of the proposed algorithm in hyperglycemia detection, achieving a 91.60\% area under the curve (AUC), 81.05\% sensitivity, and 85.54\% specificity.


\begin{thebibliography}{10}

\bibitem{CDC}
CDC, ``{National Diabetes Statistics Report}.'' \url{https://www.cdc.gov/diabetes/data/statistics-report/index.html}, 2023.
\newblock [Online; accessed 01-Jan-2024].

\bibitem{villena2019progress}
W.~Villena~Gonzales, A.~T. Mobashsher, and A.~Abbosh, ``The progress of glucose monitoring—a review of invasive to minimally and non-invasive techniques, devices and sensors,'' {\em Sensors}, vol.~19, no.~4, p.~800, 2019.

\bibitem{lazaro2020wearable}
J.~L{\'a}zaro, N.~Reljin, M.-B. Hossain, Y.~Noh, P.~Laguna, and K.~H. Chon, ``Wearable armband device for daily life electrocardiogram monitoring,'' {\em IEEE Transactions on Biomedical Engineering}, vol.~67, no.~12, pp.~3464--3473, 2020.

\bibitem{sethuraman2021mywear}
S.~C. Sethuraman, P.~Kompally, S.~P. Mohanty, and U.~Choppali, ``Mywear: a novel smart garment for automatic continuous vital monitoring,'' {\em IEEE Transactions on Consumer Electronics}, vol.~67, no.~3, pp.~214--222, 2021.

\bibitem{nguyen2012identification}
L.~L. Nguyen, S.~Su, and H.~T. Nguyen, ``Identification of hypoglycemia and hyperglycemia in type 1 diabetic patients using ecg parameters,'' in {\em 2012 Annual International Conference of the IEEE Engineering in Medicine and Biology Society}, pp.~2716--2719, IEEE, 2012.

\bibitem{nguyen2014neural}
L.~L. Nguyen, S.~Su, and H.~T. Nguyen, ``Neural network approach for non-invasive detection of hyperglycemia using electrocardiographic signals,'' in {\em 2014 36th Annual International Conference of the IEEE Engineering in Medicine and Biology Society}, pp.~4475--4478, IEEE, 2014.

\bibitem{li2021non}
J.~Li, I.~Tobore, Y.~Liu, A.~Kandwal, L.~Wang, and Z.~Nie, ``Non-invasive monitoring of three glucose ranges based on ecg by using dbscan-cnn,'' {\em IEEE journal of biomedical and health informatics}, vol.~25, no.~9, pp.~3340--3350, 2021.

\bibitem{li2023noninvasive}
J.~Li, J.~Ma, O.~M. Omisore, Y.~Liu, H.~Tang, P.~Ao, Y.~Yan, L.~Wang, and Z.~Nie, ``Noninvasive blood glucose monitoring using spatiotemporal ecg and ppg feature fusion and weight-based choquet integral multimodel approach,'' {\em IEEE Transactions on Neural Networks and Learning Systems}, 2023.

\bibitem{cordeiro2021hyperglycemia}
R.~Cordeiro, N.~Karimian, and Y.~Park, ``Hyperglycemia identification using ecg in deep learning era,'' {\em Sensors}, vol.~21, no.~18, p.~6263, 2021.

\bibitem{ingale2020ecg}
M.~Ingale, R.~Cordeiro, S.~Thentu, Y.~Park, and N.~Karimian, ``Ecg biometric authentication: A comparative analysis,'' {\em IEEE Access}, vol.~8, pp.~117853--117866, 2020.

\bibitem{woo2018cbam}
S.~Woo, J.~Park, J.-Y. Lee, and I.~S. Kweon, ``Cbam: Convolutional block attention module,'' in {\em Proceedings of the European conference on computer vision (ECCV)}, pp.~3--19, 2018.

\bibitem{adami2023contactless}
B.~Adami and N.~Karimian, ``Contactless fingerprint biometric anti-spoofing: An unsupervised deep learning approach,'' {\em arXiv preprint arXiv:2311.04148}, 2023.

\end{thebibliography}
\end{document}